\documentclass[12pt]{article}
\usepackage{amssymb,amscd,array}
\catcode `\@=11
\@addtoreset{equation}{section}

\newtheorem{thm}{Theorem}[section]

\newtheorem{lemma}[thm]{Lemma}

\def\qed{\blacksquare}
\newcommand{\be}{\begin{equation}}
\newcommand{\ee}{\end{equation}}
\newcommand{\bea}{\begin{eqnarray}}
\newcommand{\eea}{\end{eqnarray}}

\newcommand{\N}{\mathbb{N}}
\newcommand{\C}{\mathbb{C}}
\def\d{\partial}
\begin{document}
\begin{titlepage}

\begin{center}
{\bf \Large{The Interaction of Quantum Gravity with Matter\\}}
\end{center}
\vskip 1.0truecm
\centerline{D. R. Grigore, 
\footnote{e-mail: grigore@theory.nipne.ro}}
\vskip5mm
\centerline{Department of Theoretical Physics, Institute for Physics and Nuclear
Engineering ``Horia Hulubei"}
\centerline{Institute of Atomic Physics}
\centerline{Bucharest-M\u agurele, P. O. Box MG 6, ROM\^ANIA}

\vskip 2cm
\bigskip \nopagebreak
\begin{abstract}
\noindent
The interaction of (linearized) gravitation with matter is studied in the causal approach up to the second order of perturbation theory. We consider the generic case and prove that gravitation is universal in the sense that the existence of the interaction with gravitation does not put new constraints on the Lagrangian for lower spin fields. We use the formalism of quantum off-shell fields which makes our computation more straightforward and simpler.
\end{abstract}
%\newpage\setcounter{page}1
\end{titlepage}

\section{Introduction}

The general framework of perturbation theory consists in the construction of 
the chronological products such that Bogoliubov axioms are verified \cite{BS}, \cite{EG}, \cite{DF}, \cite{ano}; for every set of Wick monomials 
$ 
W_{1}(x_{1}),\dots,W_{n}(x_{n}) 
$
acting in some Fock space generated by the free fields of the model
$
{\cal H}
$
one associates the operator
$ 
T_{W_{1},\dots,W_{n}}(x_{1},\dots,x_{n}); 
$  
all these expressions are in fact distribution-valued operators called chronological products. Sometimes it is convenient to use another notation: 
$ 
T(W_{1}(x_{1}),\dots,W_{n}(x_{n})). 
$ 
The construction of the chronological products can be done recursively according to Epstein-Glaser prescription \cite{EG}, \cite{Gl} (which reduces the induction procedure to a distribution splitting of some distributions with causal support) or according to Stora prescription \cite{PS} (which reduces the renormalization procedure to the process of extension of distributions). These products are not uniquely defined but there are some natural limitation on the arbitrariness. If the arbitrariness does not grow with $n$ we have a renormalizable theory. An equivalent point of view uses retarded products \cite{St1}.

Gauge theories describe particles of higher spin. Usually such theories are not renormalizable. However, one can save renormalizability using ghost fields. Such theories are defined in a Fock space
$
{\cal H}
$
with indefinite metric, generated by physical and un-physical fields (called {\it ghost fields}). One selects the physical states assuming the existence of an operator $Q$ called {\it gauge charge} which verifies
$
Q^{2} = 0
$
and such that the {\it physical Hilbert space} is by definition
$
{\cal H}_{\rm phys} \equiv Ker(Q)/Im(Q).
$
The space
$
{\cal H}
$
is endowed with a grading (usually called {\it ghost number}) and by construction the gauge charge is raising the ghost number of a state. Moreover, the space of Wick monomials in
$
{\cal H}
$
is also endowed with a grading which follows by assigning a ghost number to every one of the free fields generating
$
{\cal H}.
$
The graded commutator
$
d_{Q}
$
of the gauge charge with any operator $A$ of fixed ghost number
\be
d_{Q}A = [Q,A]
\ee
is raising the ghost number by a unit. It means that
$
d_{Q}
$
is a co-chain operator in the space of Wick polynomials. From now on
$
[\cdot,\cdot]
$
denotes the graded commutator.
 
A gauge theory assumes also that there exists a Wick polynomial of null ghost number
$
T(x)
$
called {\it the interaction Lagrangian} such that
\be
~[Q, T] = i \partial_{\mu}T^{\mu}
\label{gauge-1}
\ee
for some other Wick polynomials
$
T^{\mu}.
$
This relation means that the expression $T$ leaves invariant the physical states, at least in the adiabatic limit. In all known models one finds out that there exists a chain of Wick polynomials
$
T^{\mu},~T^{\mu\nu},~T^{\mu\nu\rho},\dots
$
such that:
\be
~[Q, T] = i \partial_{\mu}T^{\mu}, \quad
[Q, T^{\mu}] = i \partial_{\nu}T^{\mu\nu}, \quad
[Q, T^{\mu\nu}] = i \partial_{\rho}T^{\mu\nu\rho},\dots
\label{descent}
\ee
In all cases
$
T^{\mu\nu},~T^{\mu\nu\rho},\dots
$
are completely antisymmetric in all indices; it follows that the chain of relation stops at the step $4$ (if we work in four dimensions). We can also use a compact notation
$
T^{I}
$
where $I$ is a collection of indices
$
I = [\nu_{1},\dots,\nu_{p}]~(p = 0,1,\dots,)
$
and the brackets emphasize the complete antisymmetry in these indices. All these polynomials have the same canonical dimension
\be
\omega(T^{I}) = \omega_{0},~\forall I
\ee
and because the ghost number of
$
T \equiv T^{\emptyset}
$
is supposed null, then we also have:
\be
gh(T^{I}) = |I|.
\ee
One can write compactly the relations (\ref{descent}) as follows:
\be
d_{Q}T^{I} = i~\partial_{\mu}T^{I\mu}.
\label{descent1}
\ee

For concrete models the equations (\ref{descent}) can stop earlier: for instance in the case of gravity
$
T^{\mu\nu\rho\sigma} = 0.
$
In \cite{gravity2} - \cite{ym+gravity} we have determined the most general solution of (\ref{descent}) for the case of massless gravity interacting with Yang-Mills, scalar and Dirac fields; in \cite{mass+gravity} we have considered the case of massive gravity in interaction with matter.

We can construct the chronological products
$$
T^{I_{1},\dots,I_{n}}(x_{1},\dots,x_{n}) \equiv T(T^{I_{1}}(x_{1}),\dots,T^{I_{n}}(x_{n}))
$$
according to the recursive procedure. We say that the theory is gauge invariant in all orders of the perturbation theory if the following set of identities generalizing (\ref{descent1}):
\be
d_{Q}T^{I_{1},\dots,I_{n}} = 
i \sum_{l=1}^{n} (-1)^{s_{l}} {\partial\over \partial x^{\mu}_{l}}
T^{I_{1},\dots,I_{l}\mu,\dots,I_{n}}
\label{gauge}
\ee
are true for all 
$n \in \N$
and all
$
I_{1}, \dots, I_{n}.
$
Here we have defined
\be
s_{l} \equiv \sum_{j=1}^{l-1} |I|_{j}
\ee
(see also \cite{DB}). In particular, the case
$
I_{1} = \dots = I_{n} = \emptyset
$
it is sufficient for the gauge invariance of the scattering matrix, at least
in the adiabatic limit.

Such identities can be usually broken by {\it anomalies} i.e. expressions of the type
$
A^{I_{1},\dots,I_{n}}
$
which are quasi-local and might appear in the right-hand side of the relation (\ref{gauge}). One compute these anomalies in lower orders of perturbation theory and imposing their cancellation one obtains various restrictions on the expression of the interaction Lagrangian. In this paper we consider the interaction between gravitation, Yang-Mills fields and matter (described by scalar and Fermi fields). The complete expression of the Lagrangian has been given in a previous papers \cite{gravity2} - \cite{ym+gravity}. Here we compute the anomalies of this model in the second order of perturbation theory.

We find usefull to use the formalism of off-shell fields for the computation of the anomalies. This formalism give a systematic way of computing the anomalies. Such a formalism was used previously in the literature in the context of classical field theory. We use here a pure quantum version.

In the next Section we remind our definition of free fields. We avoid explicit formulas using the reconstruction theorem of Wightmann and Borchers algebras. In Section \ref{int} we recall the main results concerning the interaction Lagrangians for various models with higher spin fields (Yang-Mills and massless gravitation). In Section \ref{off} we  introduce the off-shell formalism. Then in Section \ref{second} we describe the consequences of the cancellation of the anomalies in the second order of the perturbation theory in the most general case.
%\newpage

\section{Free Fields\label{free}}

\subsection{Free Scalar and Dirac Fields\label{scalar+dirac}}

We will adopt the description of free quantum fields given by the reconstruction theorem from axiomatic field theory \cite{J}, \cite{SW} based on Borchers algebras. In this approach one can construct a quantum field giving the Wightmann $n$-points distributions and the statistics. For a free field it is sufficient to give the Wightmann $2$-points distribution and generate the rest according to Wick theorem. We use formal distribution notations for simplicity.

A. {\bf The Real Scalar Field.} We start with the most elementary case of a real scalar field. The field is
$
\Phi(x)
$
and the Hilbert space is generated by vectors of the type
\be
\Phi(x_{1}) \cdots \Phi(x_{n})~\Omega
\label{states-scalar}
\ee
where 
$
\Omega
$
is the vacuum vector. By definition, the $2$-points distribution is
\be
<\Omega, \Phi(x_{1}) \Phi(x_{2})\Omega> = - i~D_{m}^{(+)}(x_{1} - x_{2})
\label{2-scalar}
\ee
where
$
D_{m}^{(+)}(x)
$
is the positive frequency part of the Pauli-Jordan causal distribution of mass $m$. We assume that the scalar field is a Bose field and the $n$-points distributions are generated according to Wick theorem: for $n$ odd
\be
<\Omega, \Phi(x_{1}) \cdots \Phi(x_{n})\Omega> = 0
\label{n-odd-scalar}
\ee
and for $n$ even:
\bea
<\Omega, \Phi(x_{1}) \cdots \Phi(x_{n})\Omega> = 
\sum_{\sigma} <\Omega, \Phi(x_{\sigma(1)}) \Phi(x_{\sigma(2)})\Omega> \cdots
<\Omega, \Phi(x_{\sigma(n-1)}) \Phi(x_{\sigma(n)})\Omega>;
\label{n-even-scalar}
\eea
here the sum is over all permutations $\sigma$ of the numbers
$
1, 2, \dots, n.
$
We also postulate that the field 
$
\Phi
$
is self-adjoint:
\be
\Phi^{\dagger} = \Phi.  
\label{adj-scalar}
\ee

Then one can construct the Hilbert space 
$
{\cal H}
$
from vectors of the type (\ref{states-scalar}) with the scalar product 
$
<\cdot,\cdot>
$
reconstructed from the $n$-points distributions given above and the self-adjointness assumption. We first define a sesquilinear form in the Hilbert space between two states of the form (\ref{states-scalar}) by
\bea
< \Phi(x_{n})\cdots \Phi(x_{1})\Omega, \Phi(x_{n+1})\cdots \Phi(x_{m+n})\Omega>
\nonumber \\
\equiv < \Omega, \Phi(x_{1})^{\dagger} \cdots \Phi(x_{n})^{\dagger} \Phi(x_{n+1})\cdots \Phi(x_{m+n})\Omega>
= < \Omega, \Phi(x_{1}) \cdots \Phi(x_{m+n})\Omega>
\label{scalar-product}
\eea
and one can prove that is positively defined so it induces a scalar product. 

Then the action of the scalar field on states of the form (\ref{states-scalar}) is defined in an obvious way. One can prove that the scalar field so defined verifies the Klein-Gordon equation of mass $m$
\be
K_{m} \Phi = (\square + m^{2}) \Phi = 0; \qquad 
\square \equiv \partial^{2} = \partial\cdot\partial = \partial_{\mu}~\partial^{\mu}
\label{KG-scalar-eq}
\ee
and the canonical commutations relation:
\be
[\Phi(x_{1}), \Phi(x_{2}) ] = - i~D_{m}(x_{1} - x_{2}).
\ee

Because of this commutation relation the writing of a state from the Hilbert space in the form (\ref{states-scalar}) is not unique.

Moreover, one can introduce in the Hilbert space
$
{\cal H}
$
a unitary (irreducible) representation of the Poincar\'e group according to
\bea
U_{\Lambda,a} \Phi(x) U_{\Lambda,a}^{-1} = \Phi(\Lambda^{-1}\cdot (x - a)) 
\nonumber \\
U_{\Lambda,a}~\Omega = \Omega
\label{representation-scalar}
\eea
(here 
$
\Lambda \in {\cal L}^{\uparrow}_{+}
$ 
is a proper orthochronous Lorentz transform and $a$ is a space-time translation). One can obtain in an elementary way the action of the operator 
$
U_{\Lambda,a}
$
on vectors of the type (\ref{states-scalar}) by commuting the operator with the factors
$
\Phi(x_{j})
$
till it hits the vacuum and gives the identity. In the same way one can define the space and time parity operators.

Of course, one can obtain very explicit representations for the scalar field, but they will be not needed in the following. We only mention that one can define in the same way the Wick (or normal) products
$
:\Phi^{n}(x):
$
for any integer $n$ (see \cite{WG}).

For an ensemble of real scalar fields
$
\Phi_{a}
$
where $a$ takes values in a index set $I$ we only replace (\ref{2-scalar}) by
\be
<\Omega, \Phi_{a}(x_{1}) \Phi_{b}(x_{2})\Omega> = 
- i~\delta_{ab}~D_{m}^{(+)}(x_{1} - x_{2})
\label{2-scalars}
\ee 
and we make a corresponding modification of the formula (\ref{n-even-scalar}). A complex scalar will be an apropriate combination of two real scalar fields. 

B. {\bf The Dirac Field.} For fields of spin $1/2$ the preceding formalism remains essentially the same with some technical modifications. First one needs a set of so-called Dirac matrices; they are $4 \times 4$ matrices
$
\gamma^{\mu}
$ 
verifying the Clifford algebra
\be
\{ \gamma^{\mu}, \gamma^{\nu} \} = 2~\eta^{\mu\nu}
\ee
where
$
\eta^{\mu\nu}
$
is the Poincar\'e invariant tensor (with only diagonal elements which we choose
$
1, -1, -1, -1
$). 
Such a set of matrices is essentially unique (up to some similarity transformation) according to Clifford algebra theory. In four dimension a simplified analysis due to Pauli is available (see for instance \cite{Sc1}). We also define 
\be
\gamma_{5} \equiv \gamma_{0}\gamma_{1}\gamma_{2}\gamma_{3}, \qquad
\gamma_{\epsilon} \equiv {1\over 2}~(I + \epsilon~\gamma_{5}),~\epsilon = \pm,
\qquad
\gamma^{[\mu\nu\rho]} \equiv A_{\mu\nu\rho} (\gamma^{\mu}~\gamma^{\nu}~\gamma^{\rho}).
\ee
There exists a four-dimensional representation 
$
S(A)
$
of the group 
$
SL(2,\C)
$
(the universal covering group of the proper orthochronous Lorentz group) such that
\be
S(A)^{-1} \gamma S(A) = \Lambda(A)\cdot\gamma; 
\ee
here
$
SL(2,\C) \ni A \rightarrow \Lambda(A) \in {\cal L}^{\uparrow}_{+}
$
is the covering map of the proper orthochronous Lorentz group and we note that the Lorentz group has a natural action on the Dirac matrices because they carry a Lorentz index. We will denote the matrix elements of these matrices by
$
\gamma^{\mu}_{\alpha\beta}, \alpha,\beta = 1,\cdots,4.
$

We define the causal (matrix) distribution 
\be
S_{M}(x) \equiv (i~\gamma^{\mu}~\partial_{\mu} + M)~D_{M}(x)
\ee
and the corresponding positive and negative frequency parts are
$
S_{M}^{(\pm)}.
$
Then a Dirac field is a set fields 
$
\psi_{\alpha}(x), \alpha = 1,\cdots,4
$
considered as a column ($1 \times 4$) matrix. It is convenient to define the $4 \times 1$
(line) matrix
$
\bar{\psi} = \psi^{\dagger}~\gamma_{0}.
$
Then the $2$-point distributions are given by
\bea
<\Omega, \psi_{\alpha}(x_{1}) \bar{\psi}_{\beta}(x_{2})\Omega> = 
- i~S_{M}^{(+)}(x_{1} - x_{2})_{\alpha\beta}
\nonumber \\
<\Omega, \bar{\psi}_{\alpha}(x_{1}) \psi_{\beta}(x_{2})\Omega> = 
- i~S_{M}^{(-)}(x_{2} - x_{1})_{\beta\alpha}
\label{2-dirac}
\eea
and zero in the other two remaining cases.

The $n$-points distributions are defined by a formula similar to (\ref{n-even-scalar}) but we introduce in the sum over the permutations the signature 
$
(-1)^{|\sigma|}
$
of the corresponding permutation. This will ensure the Fermi character of the Dirac field.
One can prove that the field 
$
\psi(x)
$
fulfills the Dirac equation of mass $M$: in matrix notations
\be
i~\gamma^{\mu}~\partial_{\mu} \psi(x) = M \psi(x)~\qquad\Longleftrightarrow
\qquad
i~\partial_{\mu}\bar{\psi}(x)~\gamma^{\mu} = - M~\bar{\psi}(x)
\label{Dirac-eq}
\ee
and the canonical anti-commutation relation:
\bea
\{ \psi_{\alpha}(x_{1}),~\bar{\psi}_{\beta}(x_{2}) \} = 
- i~S_{M}(x_{1} - x_{2})_{\alpha\beta}.
\label{CAR-dirac}
\eea

The unitary representation of the (proper and orthochronous) Lorentz group is defined in this case by
\bea
U_{A} \psi_{\alpha}(x) U_{A}^{-1} = 
S(A^{-1})_{\alpha\beta}~\psi_{\beta}(\Lambda(A)^{-1}\cdot x) 
\nonumber \\
U_{A}~\Omega = \Omega.
\label{representation-dirac}
\eea

One notices that there is no self-adjointness assumption on the Dirac field. There is however an spinor analogous of the real scalar field, namely the Majorana field. One can prove the existence of the so-called charge conjugation matrix $C$ defined according to
\be
\gamma_{\mu}^{T} = - C~\gamma_{\mu}~C^{-1}
\ee
i.e. the similarity transformation connecting the Dirac matrices with their transposed. Then one defines the charge conjugate spinor of $\psi$ by
\be
\psi^{C} \equiv C~\bar{\psi}^{T}
\ee
and a Majorana spinor is a spinor equal to its charge conjugate
$
\psi^{C} = \psi
$.

We close by mentioning that the choice of the statistics is crucial for obtaining a positively defined scalar product. Should we have made the opposite choice in the preceding two case, the result would have been a Fock space with a sesquilinear form, non-degenerated but not positively defined.

\subsection{Higher Spin Fields\label{yang-mills+gravity}}

For fields of higher spin one can use the preceding formalism with one major modification: it is necessary to introduce ghosts fields, which are fields with the ``wrong statistics".

C. {\bf The Massless Vector Field.} In this case we consider the vector space 
$
{\cal H}
$
of Fock type generated (in the sense of Borchers theorem) by the following fields:
$
(v^{\mu}, u, \tilde{u})
$
where the non-zero $2$-point distributions are
\bea
<\Omega, v^{\mu}(x_{1}) v^{\nu}(x_{2})\Omega> = 
i~\eta_{\mu\nu}~D_{0}^{(+)}(x_{1} - x_{2}),
\nonumber \\
<\Omega, u(x_{1}) \tilde{u}(x_{2})\Omega> = - i~D_{0}^{(+)}(x_{1} - x_{2}).
\label{2-massless-vector}
\eea

We also assume the following self-adjointness properties:
\be
v_{\mu}^{\dagger} = v_{\mu}, \qquad 
u^{\dagger} = u, \qquad
\tilde{u}^{\dagger} = - \tilde{u}.
\label{adj-vector-null}
\ee

When we generate the $n$-point function according to a formula of the type (\ref{n-odd-scalar}) and (\ref{n-even-scalar}) we assume that the field 
$
v^{\mu}
$
is Bose and the fields
$
u, \tilde{u}
$
are Fermi. When defining the unitary representation of the Lorentz group we consider that
the first field is vector and the last two are scalars. Because of the ``wrong" statistics the sesquilinear form defined by a formula of the type (\ref{scalar-product}) will not be positively defined. Nevertheless, because it is non-degenerated, we can prove that we have Klein-Gordon equations of null mass:
\be
\square~v^{\mu} = 0 \qquad \square u = 0 \qquad \square \tilde{u} = 0
\label{KG-vector-null-eq}
\ee
and the canonical commutations relation:
\bea
[v^{\mu}(x_{1}), v^{\nu}(x_{2}) ] = i~\eta^{\mu\nu}~D_{0}(x_{1} - x_{2})
\nonumber \\
\{ u(x_{1}), \tilde{u}(x_{2}) \} = - i~D_{0}(x_{1} - x_{2})
\label{CCR-vector-null}
\eea
and all other (anti)commutators are null.

We can obtain a {\it bona fid\ae} scalar product introducing the so-called {\it gauge charge} i.e. an operator $Q$ defined by:
\bea
~[Q, v^{\mu}] = i~\partial^{\mu}u,\qquad
\{ Q, u \} = 0,\qquad
\{Q, \tilde{u}\} = - i~\partial_{\mu}v^{\mu}
\nonumber \\
Q \Omega = 0.
\label{Q-vector-null}
\eea
Using these relation one can compute the action of $Q$ on any state generated by a polynomial in the fields applied on the vacuum by commuting the operator $Q$ till it hits the vacuum and gives zero. However, because of the canonical commutation relations the writing of a polynomial state is not unique. One can prove that the operator $Q$ leaves invariant the canonical (anti)commutation relations given above and this leads to the consistency of the definition. One can prove now that the operator $Q$ squares to zero:
\be
Q^{2} = 0
\ee 
and that the factor space
$
Ker(Q)/Ran(Q)
$
is isomorphic to the Fock space particles of zero mass and helicity $1$ (photons and gluons) \cite{cohomology}.

D. {\bf Massive Vector Fields.} In this case we consider the vector space 
$
{\cal H}
$
of Fock type generated by the following fields:
$
(v^{\mu}, u, \tilde{u}, \Phi)
$
where the non-zero $2$-point distributions are
\bea
<\Omega, v^{\mu}(x_{1}) v^{\nu}(x_{2})\Omega> = 
i~\eta_{\mu\nu}~D_{m}^{(+)}(x_{1} - x_{2}),
\nonumber \\
<\Omega, u(x_{1}) \tilde{u}(x_{2})\Omega> = - i~D_{m}^{(+)}(x_{1} - x_{2})
\nonumber \\
<\Omega, \Phi(x_{1}) \Phi(x_{2})\Omega> = - i~D_{m}^{(+)}(x_{1} - x_{2})
\label{2-massive-vector}
\eea
with $m$ a positive parameter. We also assume the self-adjointness properties
\be
v_{\mu}^{\dagger} = v_{\mu}, \qquad 
u^{\dagger} = u, \qquad
\tilde{u}^{\dagger} = - \tilde{u},
\qquad \Phi^{\dagger} = \Phi.
\label{adj-vector-massive}
\ee
When we generate the $n$-point function according to a formula of the type (\ref{n-odd-scalar}) and (\ref{n-even-scalar}) we assume that the fields 
$
v^{\mu}, \Phi
$
are Bose and the fields
$
u, \tilde{u}
$
are Fermi. When defining the unitary representation of the Lorentz group we consider that
the first field is vector and the last three are scalars. As in the massless case, the sesquilinear form defined by a formula of the type (\ref{scalar-product}), will not be positively defined. We can prove that we have Klein-Gordon equations of  mass $m$:
\be
(\square + m^{2})~v^{\mu} = 0 \qquad (\square + m^{2}) u = 0 \qquad 
(\square  + m^{2})\tilde{u} = 0 \qquad (\square + m^{2}) \Phi = 0
\label{KG-vector-massive-eq}
\ee
and the canonical commutations relation:
\bea
[v^{\mu}(x_{1}), v^{\nu}(x_{2}) ] = i~\eta^{\mu\nu}~D_{m}(x_{1} - x_{2})
\nonumber \\
\{ u(x_{1}), \tilde{u}(x_{2}) \} = - i~D_{m}(x_{1} - x_{2})
\nonumber \\
~[\Phi(x_{1}), \Phi(x_{2}) ] = - i~D_{m}(x_{1} - x_{2})
\label{CCR-vector-massive}
\eea
and all other (anti)commutators are null.

We can obtain a scalar product introducing the {\it gauge charge} $Q$ according to:
\bea
~[Q, v^{\mu}] = i~\partial^{\mu}u,\qquad
\{Q, u \} = 0,\qquad
\{Q, \tilde{u} \} = - i~(\partial_{\mu}v^{\mu} + m~\Phi) \qquad
[Q, \Phi ] = i~m~u
\nonumber \\
Q \Omega = 0.
\label{Q-vector-massive}
\eea
One can prove that the operator $Q$ leaves invariant the canonical (anti)commutation relations given above and this leads to the consistency of the definition. Also one can prove now that the operator $Q$ squares to zero:
\be
Q^{2} = 0
\ee 
and that the factor space
$
Ker(Q)/Ran(Q)
$
is isomorphic to the Fock space particles of positive mass and helicity $1$ (massive vector fields e.g. $W^{\pm}$ and $Z$ particles) \cite{cohomology}.

E. {\bf The (Massless) Graviton.} The Fock space is generated by the set of fields
$
(h_{\mu\nu}, u_{\rho}, \tilde{u}_{\sigma})
$
where
$
h_{\mu\nu}
$ 
is a symmetric tensor field (with Bose statistics) and
$
u_{\rho}, \tilde{u}_{\sigma}
$
are vector fields (with Fermi statistics). The (non-zero) $2$-point functions are by definition:
\bea
<\Omega, h_{\mu\nu}(x_{1}) h_{\rho\sigma}(x_{2})\Omega> = - {i\over 2}~
(\eta_{\mu\rho}~\eta_{\nu\sigma} + \eta_{\nu\rho}~\eta_{\mu\sigma}
- \eta_{\mu\nu}~\eta_{\rho\sigma})~D_{0}^{(+)}(x_{1} - x_{2}),
\nonumber \\
<\Omega, u_{\mu}(x_{1}) \tilde{u}_{\nu}(x_{2})\Omega> = i~\eta_{\mu\nu}~
D_{0}^{(+)}(x_{1} - x_{2})
\label{2-graviton-null}
\eea
and we assume also that
\be
h_{\mu\nu}^{\dagger} = h_{\mu\nu}, \qquad 
u_{\rho}^{\dagger} = u_{\rho}, \qquad
\tilde{u}_{\sigma}^{\dagger} = - \tilde{u}_{\sigma}.
\label{adj-graviton-null}
\ee

One can prove the validity of the equations of motion
\be
\square~h_{\mu\nu} = 0 \qquad \square~u_{\rho} = 0 \qquad 
\square\tilde{u}_{\sigma} = 0
\label{KG-graviton-null-eq}
\ee
and the canonical (anti)commutation relations
\bea
[ h_{\mu\nu}(x_{1}), h_{\rho\sigma}(x_{2}) ] = - {i\over 2}~
(\eta_{\mu\rho}~\eta_{\nu\sigma} + \eta_{\nu\rho}~\eta_{\mu\sigma}
- \eta_{\mu\nu}~\eta_{\rho\sigma})~D_{0}(x_{1} - x_{2}),
\nonumber \\
\{ u_{\mu}(x_{1}), \tilde{u}_{\nu}(x_{2}) \} = i~\eta_{\mu\nu}~D_{0}(x_{1} - x_{2})
\label{CCR-graviton-null}
\eea
and all other (anti)commutators are null. One defines the gauge charge according to
\be
~[Q, h_{\mu\nu}] = - {i\over 2}~(\partial_{\mu}u_{\nu} + \partial_{\nu}u_{\mu}
- \eta_{\mu\nu} \partial_{\rho}u^{\rho}),\qquad
\{Q, u_{\mu}\} = 0,\qquad
\{Q, \tilde{u}_{\mu}\} = i~\partial^{\nu}h_{\mu\nu},
\label{Q-graviton-null}
\ee
proves that is well defined and that the factor space 
$
Ker(Q)/Ran(Q)
$
is isomorphic to the Fock space particles of zero mass and helicity $2$ (gravitons) \cite{gravity2}. In the same way one can treat the case of massive gravity. 

F. We can combine the cases C, D and A described above as follows. We take 
$
I = I_{1} \cup I_{2} \cup I_{3}
$
a set of indices and we make the following choices:

(a) for
$
a \in I_{1}
$
we consider the set of fields
$
(v^{\mu}_{a}, u_{a}, \tilde{u}_{a})
$
of null mass;

(b) for 
$
a \in I_{2}
$
we consider the fields
$
(v^{\mu}_{a}, u_{a}, \tilde{u}_{a},\Phi_{a})
$
of positive mass 
$
m_{a};
$

(c) for 
$
a \in I_{3}
$
we consider the scalar fields
$
\Phi_{a}
$
of mass
$
m_{a}^{H};
$
they are called {\it Higgs fields}. We define the mass matrix $m$ according to
\be
m_{ab} \equiv \delta_{ab}~m_{a},~~a, b \in I_{2} \cup I_{3}.
\label{mass-ym}
\ee
The expressions for the $2$-points distributions are easily given if we want that the final Fock should be a tensor space of
$
|I_{1}|
$ 
copies of massless vector fields,
$
|I_{2}|
$ 
copies of massive vectors fields and
$
|I_{3}|
$ 
copies of scalar fields. Equivalently, we can consider that we have the fields
$
(v^{\mu}_{a}, u_{a}, \tilde{u}_{a},\Phi_{a})
$
for all indices
$
a \in I_{1} \cup I_{2} \cup I_{3}
$
but we have
$
\Phi_{a},~a \in I_{1}
$
and
$
v^{\mu}_{a} =0, u_{a} = 0, \tilde{u}_{a} = 0,~a \in I_{3}.
$
In the most general case we add cases B and E: 

(d) for 
$
A \in I_{4}
$
(another index set) we consider the Dirac fields 
$
\psi_{A}.
$
As above we define the mass matrix of the Dirac fields by
\be
M_{AB} \equiv \delta_{AB}~M_{A},~~A, B \in I_{4}.
\label{mass-dirac}
\ee

(e) the (massless) gravitons.

The matter is described by the Dirac fields together with the Higgs fields.
\section{Interactions\label{int}}

The discussion from the Introduction provides the physical justification for determining the cohomology of the operator 
$
d_{Q} = [Q,\cdot]
$
(where by 
$
[\cdot,\cdot]
$
we always mean the graded commutator) induced by $Q$ in the space of Wick polynomials. A polynomial 
$
p \in {\cal P}
$
verifying the relation
\be
d_{Q}p = i~\d_{\mu}p^{\mu}
\label{rel-co}
\ee
for some polynomials
$
p^{\mu}
$
is called a {\it relative co cycle} for 
$
d_{Q}.
$
The expressions of the type
\be
p = d_{Q}b + i~\d_{\mu}b^{\mu}, \qquad (b, b^{\mu} \in {\cal P})
\ee
are relative co-cycles and are called {\it relative co-boundaries}. We denote by
$
Z_{Q}^{\rm rel}, B_{Q}^{\rm rel} 
$
and
$
H_{Q}^{\rm rel}
$
the corresponding cohomological spaces. In (\ref{rel-co}) the expressions
$
p_{\mu}
$
are not unique. It is possible to choose them Lorentz covariant. We have a general description of the most general form of the interaction of the previous fields which combines results from \cite{cohomology} and \cite{gravity2}. Summation over the dummy indices is used everywhere. For simplicity we do not write the double dots of the Wick product notations.

A. {\bf Yang-Mills Fields}

Here we combine cases C, D, A and B.
\begin{thm}
Let $T$ be a relative co-cycle in the variables 
$
(v^{\mu}_{a}, u_{a}, \tilde{u}_{a},\Phi_{a}),~a \in I_{1} \cup I_{2} \cup I_{3}
$
and
$
\psi_{A},~A \in I_{4}
$
which is tri-linear in the fields, of canonical dimension
$
\omega(T) \leq 4
$
and ghost number
$
gh(T) = 0.
$
Then:
(i) $T$ is (relatively) cohomologous to a non-trivial co-cycle of the form:
\bea
t_{\rm ym} = f_{abc} \left( {1\over 2}~v_{a\mu}~v_{b\nu}~F_{c}^{\nu\mu}
+ u_{a}~v_{b}^{\mu}~\d_{\mu}\tilde{u}_{c}\right)
\nonumber \\
+ f^{\prime}_{abc} (\Phi_{a}~\phi_{b}^{\mu}~v_{c\mu} + m_{b}~\Phi_{a}~\tilde{u}_{b}~u_{c})
\nonumber \\
+ {1\over 3!}~f^{\prime\prime}_{abc}~\Phi_{a}~\Phi_{b}~\Phi_{c}
+ \sum_{\epsilon = \pm}~
\overline{\psi} t^{\epsilon}_{a} \otimes \gamma^{\mu}\gamma_{\epsilon} \psi~v_{a\mu} 
+ \sum_{\epsilon = \pm}~
\overline{\psi} s^{\epsilon}_{a} \otimes \gamma_{\epsilon} \psi~\Phi_{a}
\label{int-ym}
\eea

(ii) The relation 
$
d_{Q}t_{\rm ym} = i~\d_{\mu}t^{\mu}_{\rm ym}
$
is verified by:
\be
t_{\rm ym}^{\mu} = f_{abc} \left( u_{a}~v_{b\nu}~F^{\nu\mu}_{c} -
{1\over 2} u_{a}~u_{b}~\d^{\mu}\tilde{u}_{c} \right)
+ f^{\prime}_{abc}~\Phi_{a}~\phi_{b}^{\mu}~u_{c}
+ \sum_{\epsilon = \pm}~
\overline{\psi} t^{\epsilon}_{a} \otimes \gamma^{\mu}\gamma_{\epsilon} \psi~u_{a}
\label{t-mu-ym}
\ee

(iii) The relation 
$
d_{Q}t^{\mu}_{\rm ym} = i~\d_{\nu}t^{\mu\nu}_{\rm ym}
$
is verified by
\be
t_{\rm ym}^{\mu\nu} \equiv {1\over 2} f_{abc}~u_{a}~u_{b}~F_{c}^{\mu\nu}.
\label{t-munu-ym}
\ee
and we have
$
d_{Q}t^{\mu\nu}_{\rm ym} = 0.
$

(iv) The expressions given above are self-adjoint iff the constants
$
f_{abc},~f^{\prime}_{abc}
$
and
$
f^{\prime\prime}_{abc}
$
are real and the matrices
$
t^{\epsilon}_{a},~i~s^{\epsilon}_{a}
$
are self-adjoint.
\label{t-ym}
\end{thm}

Here we have defined the gauge invariants which are not coboundaries
\be
F^{\mu\nu}_{a} \equiv \d^{\mu}v^{\nu}_{a} - \d^{\nu}v^{\mu}_{a}, 
\quad \forall a \in I_{1} \cup I_{2}
\ee 
and
\be
\phi^{\mu}_{a} \equiv \d^{\mu}\Phi_{a} - m_{a}~v^{\mu}_{a}, 
\quad \forall a \in I_{2}
\ee

and the various constants are constrained as follows:

(a) first, we have some limitations on the range of the indices because of the definitions of the sets of indices 
$
I_{j},~j= 1,2,3
$
namely: (a1)
$
f_{abc} = 0
$
if one of the indices is in
$
I_{3};
$
(a2) also
$
f^{\prime}_{abc} = 0
$
if 
$
c \in I_{3}
$
or one of the indices $a$ and $b$ are from
$
I_{1};
$
(a3) the matrices
$
t^{\epsilon}_{a}
$
are zero if
$
a \in I_{3};
$
(a4) the matrices
$
s^{\epsilon}_{a}
$
are zero if
$
a \in I_{1}.
$

(b) The constants
$
f_{abc}
$
are completely antisymmetric
\be
f_{abc} = f_{[abc]}.
\label{anti-f}
\ee

(c) The expressions
$
f^{\prime}_{abc}
$
are antisymmetric  in the indices $a$ and $b$:
\be
f^{\prime}_{abc} = - f^{\prime}_{bac}
\label{anti-f'}
\ee
and are connected to 
$f_{abc}$
by:
\be
f_{abc}~m_{c} = f^{\prime}_{cab} m_{a} - f^{\prime}_{cba} m_{b}.
\label{f-f'}
\ee

(d) The (completely symmetric) expressions 
$f^{\prime\prime}_{abc} = f^{\prime\prime}_{\{abc\}}$
verify
\be
f^{\prime\prime}_{abc} = \left\{\begin{array}{rcl} 
{1 \over m_{c}}~f'_{abc}~(m_{a}^{2} - m_{b}^{2}) & \mbox{for} & a, b \in I_{3}, c \in I_{2} \\
- {1 \over m_{c}}~f'_{abc}~m_{b}^{2} & \mbox{for} & a, c \in I_{2}, b \in I_{3}.\end{array}\right.
\label{f"}
\ee

(e) The matrices 
$
t^{\epsilon}_{a}
$
and
$
s^{\epsilon}_{a}
$
are connected by:
\be
m_{a}~s_{a}^{\epsilon} = i(M~t^{\epsilon}_{a} - t^{-\epsilon}_{a}~M).
\label{conserved-current}
\ee

B. {\bf Gravitation}

\begin{thm}
Let $T$ be a relative co-cycle in the variables
$
(h_{\mu\nu}, u_{\rho}, \tilde{u}_{\sigma})
$
which is tri-linear in the fields, of canonical dimension
$
\omega(T) \leq 5
$
and ghost number
$
gh(T) = 0.
$
Then:
(i) $T$ is (relatively) cohomologous to a non-trivial co-cycle of the form:
\bea
t_{\rm gr} = \kappa ( 2~h_{\mu\rho}~\d^{\mu}h^{\nu\lambda}~\d^{\rho}h_{\nu\lambda}
+ 4~h_{\nu\rho}~\d_{\lambda}h^{\mu\nu}~\d_{\mu}h^{\rho\lambda}
- 4~h_{\rho\lambda}~\d^{\mu}h^{\nu\rho}~\d_{\mu}{h_{\nu}}^{\lambda}
\nonumber \\
+ 2~h^{\rho\lambda}~\d_{\mu}h_{\rho\lambda}~\d^{\mu}h
- h_{\mu\rho}~\d^{\mu}h~\d^{\rho}h 
- 4~u^{\rho}~\d^{\nu}\tilde{u}^{\lambda}~\d_{\rho}h_{\nu\lambda}
\nonumber \\
+ 4~\d^{\rho}u^{\nu}~\d_{\nu}\tilde{u}^{\lambda}~h_{\rho\lambda}
+ 4~\d^{\rho}u^{\nu}~\d^{\lambda}\tilde{u}_{\nu}~h_{\rho\lambda}
- 4~\d^{\nu}u_{\nu}~\d^{\rho}\tilde{u}^{\lambda}~h_{\rho\lambda})
\eea

(ii) The relation 
$
d_{Q}t_{\rm gr} = i~\d_{\mu}t^{\mu}_{\rm gr}
$
is verified by:
\bea
t^{\mu}_{\rm gr} = \kappa ( - 2 u^{\mu}~\d_{\nu}h_{\rho\lambda}~\d^{\rho}h^{\nu\lambda} 
+ u^{\mu}~\d_{\rho}h_{\nu\lambda}~\d^{\rho}h^{\nu\lambda} 
- {1\over 2} u^{\mu}~\d_{\rho}h~\d^{\rho}h
\nonumber \\
+ 4~u^{\rho}~\d^{\nu}h^{\mu\lambda}~\d_{\rho}h_{\nu\lambda}
- 2~u^{\rho}~\d^{\mu}h^{\nu\lambda}~\d_{\rho}h_{\nu\lambda}
+ u^{\rho}~\d^{\mu}h~\d_{\rho}h
\nonumber \\
- 4~\d^{\rho}u^{\nu}~\d_{\nu}h^{\mu\lambda}~h_{\rho\lambda}
- 4~\d^{\rho}u_{\nu}~\d^{\lambda}h^{\mu\nu}~h_{\rho\lambda}
+ 4~\d^{\lambda}u_{\rho}~\d^{\mu}h^{\nu\rho}~h_{\nu\lambda}
\nonumber \\
+ 4~\d_{\nu}u^{\nu}~\d^{\rho}h^{\mu\lambda}~h_{\rho\lambda}
- 2~\d_{\nu}u^{\nu}~\d^{\mu}h^{\rho\lambda}~h_{\rho\lambda}
- 2~\d^{\rho}u^{\lambda}~h_{\rho\lambda}~\d^{\mu}h
+ \d^{\nu}u_{\nu}~h~\d^{\mu}h
\nonumber \\
- 2~u^{\mu}~\d_{\nu}\d_{\rho}u^{\rho}~\tilde{u}^{\nu}
+ 2~u_{\rho}~\d^{\rho}\d^{\sigma}u_{\sigma}~\tilde{u}^{\mu}
- 2~u^{\mu}~\d_{\lambda}u_{\rho}~\d^{\rho}\tilde{u}^{\nu}
\nonumber \\
+ 2~u_{\rho}~\d_{\lambda}u^{\mu}~\d^{\rho}\tilde{u}^{\lambda}
+ 2~\d^{\rho}u_{\rho}~\d_{\lambda}u^{\mu}~\tilde{u}^{\lambda}
- 2~u_{\rho}~\d^{\rho}u_{\lambda}~\d^{\mu}\tilde{u}^{\lambda})
\eea

(iii) The relation 
$
d_{Q}t^{\mu}_{\rm gr} = i~\d_{\nu}t^{\mu\nu}_{\rm gr}
$
is verified by:
\bea
t^{\mu\nu}_{\rm gr} \equiv \kappa 
[ 2 ( - u^{\mu}~\d_{\lambda}u_{\rho}~\d^{\rho}h^{\nu\lambda}
+ u_{\rho}~\d_{\lambda}u^{\mu}~\d^{\rho}h^{\nu\lambda}
+ u_{\rho}~\d^{\rho}u_{\lambda}~\d^{\nu}h^{\mu\lambda}
+ \d_{\rho}u^{\rho}~\d_{\lambda}u^{\mu}~h^{\nu\lambda})
\nonumber \\
- (\mu \leftrightarrow \nu)
+ 4~\d^{\lambda}u^{\mu}~\d^{\rho}u^{\nu}~h_{\rho\lambda} ].
\eea

(iv) The relation 
$
d_{Q}t^{\mu\nu}_{\rm gr} = i~\d_{\rho}t^{\mu\nu\rho}_{\rm gr}
$
is verified by:
\bea
t^{\mu\nu\rho}_{\rm gr} \equiv 
\kappa [ 2 u_{\lambda}~\d^{\lambda}u^{\rho}~u^{\mu\nu}
- u_{\rho}~(\d^{\mu}u^{\lambda}~\d_{\lambda}u^{\nu}
- \d^{\nu}u^{\lambda}~\d_{\lambda}u^{\mu}) 
+ {\rm circular~perm.}]
\eea
and we have
$
d_{Q}t^{\mu\nu\rho}_{\rm gr} = 0.
$

(v) The preceding expressions are self-adjoint iff the constant $\kappa$ is real.
\label{T-gr}
\end{thm}
The gauge invariants which are not coboundaries are more complicated to write. Beside
$
u_{\mu}
$
and the antisymmetric combination
\be
u_{[\mu\nu]} \equiv {1\over 2}~(\d_{\mu}u_{\nu} - \d_{\nu}u_{\mu})
\ee
we also have the linear part of the Riemann tensor from which we must subtract all traces \cite{gravity2}.

C. {\bf Interaction between Yang-Mills Fields and Gravity} 
\begin{thm}
Let $T$ be a relative co-cycle depending on the Yang-Mills and gravitational variables, 
which is tri-linear in the fields, of canonical dimension
$
\omega(T) \leq 5
$
and ghost number
$
gh(T) = 0.
$
Then:
(i) $T$ is (relatively) cohomologous to a non-trivial co-cycle of the form:
\bea
t_{\rm int} \equiv ~f_{ab}~(4 h_{\mu\nu}~F_{a}^{\mu\rho}~{F_{b}^{\nu}}_{\rho} 
- h~F_{a\rho\sigma}~F_{b}^{\rho\sigma} 
+ 4~u_{\mu}~d_{\nu}\tilde{u}_{a}~F_{b}^{\mu\nu}
- h_{\mu\nu}~\phi_{a}^{\mu}~\phi_{b}^{\nu}
+ m_{a}~u_{\mu}~\tilde{u}_{a}~\phi_{b}^{\mu})
\nonumber \\
+ f^{\prime}_{cd}~\left(h_{\mu\nu}~\d^{\mu}\Phi_{c}~\d^{\nu}\Phi_{d} 
- {m^{2}_{a} + m^{2}_{b} \over 4}~h~\Phi_{a}~\Phi_{b}\right)
\nonumber \\
+ \left(h_{\mu\nu} - {1\over 2}~\eta_{\mu\nu}~h\right)~
(\d^{\mu}\bar{\psi}~c^{\epsilon} \otimes \gamma^{\nu}\gamma_{\epsilon}\psi
- \bar{\psi}~c^{\epsilon} \otimes \gamma^{\nu}\gamma_{\epsilon}\d^{\mu}\psi).
\eea

(ii) The relation 
$
d_{Q}t_{\rm int} = i~\d_{\mu}t^{\mu}_{\rm int}
$
is verified by:
\bea
t_{\rm int}^{\mu} \equiv f_{ab}~( u^{\mu}~F_{a}^{\rho\sigma}~F_{b\rho\sigma} 
+ 4~u^{\rho}~F_{a}^{\mu\nu}~F_{b\nu\rho} 
- 2~u^{\mu}~\phi_{a}^{\nu}~\phi_{b\nu}
+ 4~u_{\nu}~\phi_{a}^{\mu}~\phi_{b}^{\nu})
\nonumber \\
+ f^{\prime}_{cd}~\left({1\over 2}~u^{\mu}~\d_{\nu}\Phi_{c}~\d^{\nu}\Phi_{d} 
- u_{\nu}~\d^{\mu}\Phi_{c}~\d^{\nu}\Phi_{d} 
- {m^{2}_{a} + m^{2}_{b} \over 4}~u^{\mu}~\Phi_{a}~\Phi_{b}\right)
\nonumber \\
- {1\over 2}~u_{\nu}~
[ (\d^{\mu}\bar{\psi}~c^{\epsilon} \otimes \gamma^{\nu}\gamma_{\epsilon}\psi
- \bar{\psi}~c^{\epsilon} \otimes \gamma^{\nu}\gamma_{\epsilon}\d^{\mu}\psi)
+ (\mu \leftrightarrow \nu) ]
\eea
and we also have
\be
d_{Q}t_{\rm int}^{\mu} = 0.
\ee

(ii) The constants
$
f_{ab}
$
are non-zero only for 
$
a, b \in I_{1} \cup I_{2}
$
and
$
f^{\prime}_{cd}
$
are non-zero only for
$
c, d \in I_{3};
$
these two matrices are symmetric and commute with the mass matrix:
\be
f_{ab}~(m_{a} - m_{b}) = 0,\qquad
f^{\prime}_{cd}(m_{c} - m_{d}) = 0.
\ee
The matrices
$
c^{\epsilon}
$
verify the identity:
\be
M~c^{\epsilon} = c^{- \epsilon}~M
\ee

(iii) The preceding expressions are self-adjoint iff the constants  
$
f_{ab}
$
and
$
f^{\prime}_{cd}
$
are real and we also have
\be
(c^{\epsilon})^{\dagger} = - c^{- \epsilon}.
\ee
\label{T-int}
\end{thm}

In all three theorems the co-cycles
$
t^{I}
$
are non-trivial. 

There are different ways to obtain the preceding results. One can proceed by brute force, making an ansatz for the expressions
$
T^{I}
$
and solving the identities of the type (\ref{descent1}) as it is done in \cite{Sc2}. There are some tricks to simplify such a computation. The first one makes an ansatz for 
$
T
$
and eliminates the most general relative cocycle. Then one computes 
$
d_{Q}T
$
and writes it as a total divergence plus terms without derivatives on the ghost fields. Another trick is to use the so-called descent procedure. We present them in the most elementary case of pure Yang-Mills fields (only the set 
$
I_{1}
$
is non-void). We have the general form:
\bea
T = f^{(1)}_{abc} v_{a}^{\mu} v_{b}^{\nu} \partial_{\mu}v_{c\mu} 
+ f^{(2)}_{abc} v_{a}^{\mu} v_{b\mu} \partial_{\nu}v_{c}^{\nu} 
+ f^{(3)}_{abc}~\epsilon_{\mu\nu\rho\sigma}~v_{a}^{\mu} v_{b}^{\nu} \partial^{\sigma}v_{c}^{\rho}
\nonumber \\
+ g^{(1)}_{abc} v^{\mu}_{a} u_{b} \partial_{\mu}\tilde{u}_{c}
+ g^{(2)}_{abc} \partial_{\mu}v^{\mu}_{a} u_{b} \tilde{u}_{c}
+ g^{(3)}_{abc} v^{\mu}_{a} \partial_{\mu}u_{b} \tilde{u}_{c}.
\eea
Eliminating relative coboundaries we can fix: 
\be
f^{(1)}_{abc} = - f^{(1)}_{bac},\qquad
f^{(2)}_{abc} = 0,\qquad
g^{(3)}_{abc} = 0,\qquad
g^{(2)}_{abc} = g^{(2)}_{bac}.
\ee
Then we obtain easily:
\be
d_{Q}T = i u_{a} T_{a}  + {\rm total~div}
\ee
where:
\bea
T_{a} = - 2 f^{(1)}_{abc}~\partial^{\nu}v^{\mu}_{b}~\partial_{\mu}v_{c\nu}
+ (f^{(1)}_{cba} + g^{(2)}_{bac})~
\partial_{\mu}v^{\mu}_{b}~\partial_{\nu}v_{c}^{\nu}
\nonumber \\
+ (- f^{(1)}_{abc} + f^{(1)}_{cba} + f^{(1)}_{bca} + g^{(1)}_{bca})~
v^{\mu}_{b}~\partial_{\mu}\partial_{\nu}v^{\nu}_{c}
\nonumber \\
- 2 f^{(3)}_{abc}~\epsilon_{\mu\nu\rho\sigma}
~\partial^{\mu}v^{\nu}_{b}~\partial_{\sigma}v_{c\rho}.
\eea

Now the gauge invariance condition (\ref{gauge-1}) becomes
\be
u_{a} T_{a} = \partial_{\mu}t^{\mu}
\label{gauge-11}
\ee
for some expression
$
t^{\mu}
$
which has, from power counting arguments, the general form
\be
t^{\mu} = u_{a}~t^{\mu}_{a} + \partial^{\mu}u_{a}~t_{a}
+ \partial_{\nu}u_{a}~t^{\mu\nu}_{a}
\ee
where the polynomial
$
t^{\mu\nu}_{a}
$
does not contain terms with the factor 
$
\eta^{\mu\nu}.
$
Then the relation (\ref{gauge-11}) is equivalent to:
\bea
\d_{\mu}t^{\mu}_{a} - m_{a}^{2}~t_{a} = T_{a}
\nonumber \\
t^{\mu}_{a} + \d^{\mu}t_{a} + \d_{\nu}t^{\nu\mu}_{a} = 0
\nonumber \\
t^{\mu\nu}_{a} = t^{\nu\mu}_{a}.
\eea
One can obtain easily from this system that
\be
T_{a} = (\square + m_{a}^{2})~t_{a}.
\ee
Writing a generic form for 
$
t_{a}
$
it is easy to prove that in fact:
\be
T_{a} = 0;
\ee
from here we easily obtain the total antisymmetry of the expressions
$
f^{(1)}_{abc}
$
and
$
f^{(3)}_{abc};
$
also we have
$
g^{(2)}_{abc} = 0.
$
Now one can take 
$
f^{(3)}_{abc} = 0
$
if we subtract from $T$ a total divergence. As a result we obtain the (unique) solution:
\be
T = f^{(1)}_{abc} ( v_{a}^{\mu} v_{b}^{\nu} \partial_{\nu}v_{c\mu}
- v_{a}^{\mu} u_{b} \partial_{\mu}\tilde{u}_{c})
\ee
which is the expression from the first theorem in this particular case i.e.the first line of (\ref{int-ym}).

Now we briefly present the descent method in this case. There are two results which must be used repeatedly \cite{cohomology}. First, we have a version of the Poincar\'e lemma valid for Wick monomials and then we have a description of the cohomology group
$
H_{Q}
$
of
$
d_{Q}
$
in terms of invariants: if $T$ is a Wick polynomial verifying
$
d_{Q}~T = 0
$
then it is of the form 
$
T = d_{Q}B + T_{0}
$
where 
$
T_{0}
$
depends only on the gauge invariants
$
u_{a}, F_{a}^{\mu\nu}.
$

By hypothesis we have
\be
d_{Q}T = i~\d_{\mu}T^{\mu}.
\label{descent-a}
\ee
If we apply 
$
d_{Q}
$
we obtain
$
\d_{\mu}d_{Q}~T^{\mu} = 0.
$
Using Poincar\'e lemma so one finds out some Wick polynomials
$
T^{[\mu\nu]}
$
such that
\be
d_{Q}T^{\mu} = i~\d_{\nu}T^{[\mu\nu]}.
\label{descent-b}
\ee
Continuing in the same way we find
$
T^{[\mu\nu\rho]}
$
such that
\be
d_{Q}T^{[\mu\nu]} = i~\d_{\rho}T^{[\mu\nu\rho]};
\label{descent-c}
\ee
we also have
\be
gh(T^{I}) = |I|.
\ee 

It means that
$
T^{[\mu\nu\rho]}
$
is a sum of terms of the type
$
\eta^{\mu\nu}~u_{a}~u_{b}~\d^{\rho}u_{c}
$
i.e. is a coboundary:
$
T^{[\mu\nu\rho]} = d_{Q}B^{[\mu\nu\rho]}.
$
We introduce in (\ref{descent-c}) and obtain 
\be
d_{Q}(T^{[\mu\nu]} - i~\d_{\rho}B^{[\mu\nu\rho]}) = 0.
\ee
Using the description of the cohomology of
$
d_{Q}
$
we can easily find that we have:
\be
T^{[\mu\nu]} = d_{Q}B^{[\mu\nu]}+ i~\d_{\rho}B^{[\mu\nu\rho]} + T^{[\mu\nu]}_{0}
\ee 
where the last term depends only on the invariants 
$
u_{a}, F_{a}^{\mu\nu}
$
i.e.
\be
T_{0}^{[\mu\nu]} = {1\over 2}~f^{(1)}_{[ab]c}~u_{a}~u_{b}~F_{c}^{\mu\nu}
+ {1\over 2}~f^{(2)}_{[ab]c}~\epsilon^{\mu\nu\rho\sigma}~u_{a}~u_{b}~F_{c\rho\sigma};
\ee
We substitute these expressions in (\ref{descent-b}) and obtain 
\be
d_{Q}(T^{\mu} - i~\d_{\nu}B^{[\mu\nu]} - t^{\mu}) = 0
\ee
where:
\be
t^{\mu} \equiv f^{(1)}_{[ab]c}~\left(u_{a}~v_{b\nu}~F_{c}^{\nu\mu}
- {1\over 2}~u_{a}~u_{b}~\d^{\mu}\tilde{u}_{c}\right)
- f^{(2)}_{[ab]c}~\epsilon^{\mu\nu\rho\sigma}~u_{a}~v_{b\nu}~F_{c\rho\sigma}.
\ee

If we use again the cohomology of
$
d_{Q}
$
we can easily find out that in fact:
\be
T^{\mu} = d_{Q}B^{\mu} + i~\d_{\nu}B^{[\mu\nu]} + t^{\mu}.
\ee
We substitute this in (\ref{descent-a}) and we obtain the restrictions
\be
f^{(1)}_{[ab]c} = - f^{(1)}_{[ac]b}, \qquad f^{(2)}_{[ab]c} = - f^{(2)}_{[ac]b}
\nonumber
\ee
so the constants 
$
f^{(1)}_{[ab]c},~f^{(2)}_{[ab]c}
$
are in fact completely antisymmetric and
\be
d_{Q}(T^{\mu} - i~\d_{\nu}B^{\mu\nu} - t) = 0
\ee
where
\be
t \equiv f^{(1)}_{[abc]} \left( {1\over 2}~v_{a\mu}~v_{b\nu}~F_{c}^{\nu\mu}
+ u_{a}~v_{b}^{\mu}~\d_{\mu}\tilde{u}_{c}\right)
- {1\over  2}~f^{(2)}_{[abc]}~
\epsilon_{\mu\nu\rho\sigma}~v_{a}^{\mu}~v^{\nu}_{b}~F^{\rho\sigma}_{c}. 
\ee
The description of the cohomology of
$
d_{Q}
$
leads to
\be
T = d_{Q}B + i~\d_{\mu}B^{\mu} + t.
\ee
Finally one proves that the last term from the expression $t$ is a total divergence.

In the general case one can use both methods also. In the descent method we will have more gauge invariants in the description of the cohomology group
$
H_{Q}
$.
\newpage 
\section{The Off-Shell Formalism\label{off}}

It is known that in the second order of the perturbation theory some anomalies can appear and this is due essentially because the Pauli-Jordan distribution 
$
D_{m}
$
verifies Klein-Gordon equation:
\be
K_{m}~D_{m} = (\square + m^{2})~D_{m} = 0.
\label{kg-d}
\ee
but the associated Feynman distribution
$
D_{m}^{F}
$
verifies
\be
K_{m}~D_{m}^{F} = (\square + m^{2})~D_{m} = \delta(x - y).
\label{kg-df}
\ee

Let us describe in detail this point. One computes the second order causal commutator and finds out that the tree contribution has the following generic form:
\bea
[T^{I_{1}}(x),T^{I_{2}}(y)]^{\rm tree} = 
\sum_{m}~[~D_{m}(x - y)~A^{I_{1},I_{2}}_{m}(x,y) 
+ \d_{\alpha}~D_{m}(x - y)~A^{I_{1},I_{2};\alpha}_{m}(x,y) 
\nonumber \\
+ \d_{\alpha}\d_{\beta}~D_{m}(x - y)~A^{I_{1},I_{2};\{\alpha\beta\}}_{m}(x,y)~]
\eea
where the sum runs over the various masses from the spectum of the model and the expressions
$
A^{I_{1},I_{2}}_{m}, A^{I_{1},I_{2};\alpha}_{m}, A^{I_{1},I_{2};\{\alpha\beta\}}_{m}
$
are Wick polynomials. Moreover we have from (\ref{descent1}) the identity
\be
d_{Q}[T^{I_{1}}(x),T^{I_{2}}(y)] = i {\d \over \d x^{\mu}}[T^{I_{1}\mu}(x),T^{I_{2}}(y)]
+ (-1)^{|I_{1}|} {\d \over \d y^{\mu}}[T^{I_{1}}(x),T^{I_{2}\mu}(y)]
\label{d-2}
\ee
which stays true if we take only the tree graphs. Now one can find out the corresponding chronological products by simply substituting in the preceding expression the causal distribution by the associated Feynman propagator:
$
D_{m} \rightarrow D_{m}^{F}
$
i.e.
\bea
T^{I_{1},I_{2}}(x,y)^{\rm tree} = 
\sum_{m}~[~D_{m}^{F}(x - y)~A^{I_{1},I_{2}}_{m}(x,y) 
+ \d_{\alpha}~D_{m}^{F}(x - y)~A^{I_{1},I_{2};\alpha}_{m}(x,y) 
\nonumber \\
+ \d_{\alpha}\d_{\beta}~D_{m}^{F}(x - y)~A^{I_{1},I_{2};\{\alpha\beta\}}_{m}(x,y)~]
\eea
In this way all Bogoliubov axioms are true (in the second order) but we might break gauge invariance i.e. the identity (\ref{gauge}) for 
$
n = 2
$
\be
d_{Q}T^{I_{1},I_{2}}(x,y) = i {\d \over \d x^{\mu}}T^{I_{1}\mu,I_{2}}(x,y)
+ (-1)^{|I_{1}|} {\d \over \d y^{\mu}}T^{I_{1},I_{2}\mu}(x,y)
\label{t-2}
\ee
might not be true. Indeed, let us consider the simplest case 
$
I_{1} = [\mu], I_{2} = \emptyset
$
and suppose that in the chronological product 
$
T(T^{\mu}(x),T(y))^{\rm tree}
$
we have a term of the type
$
\d^{\mu}D_{m}^{F}(x - y)~A(x,y).
$ 

Then, because of the difference between the relations (\ref{kg-d}) and (\ref{kg-df}) we have in the right hand side of (\ref{t-2}) an extra-term 
$
2~\delta(x - y)~A(x,y).
$
One must collect all quasi-local terms appearing in this way and check if they can be put under the form 
$
d_{Q}R(x,y) + i {\d \over \d x^{\mu}}R^{\mu}(x,y) + {\d \over \d y^{\mu}}R^{\mu}(y,x).
$ 
If this can be done, then we can restore gauge invariance (at least for the tree contributions) by redefining the chronological products in an obvious way. 

So the first problem is to find out the anomaly i.e. the expression appearing in the right hand side of (\ref{t-2}) and the second problem is to see in which conditions it can be eliminated by a redefinition of the chronological products. Even the first problem is not exactly elementary in complex models as for instance the case of gravity; in \cite{Sc2} one can see for instance that not only terms of the type 
$
\d^{\mu}D_{m}^{F}(x - y)~A(x,y)
$ 
can produce anomalies. So we need a systematic way to compute the anomaly.

This suggests to make the following change in the description of the fields from section \ref{free}, namely we replace the Pauli-Jordan distribution 
$
D_{m}
$
by some off-shell distribution 
$
D_{m}^{\rm off}
$
which does not verify Klein-Gordon equation but converges in some limit (in the sense of distribution theory) to
$
D_{m}.
$ 
For instance we can take 
\be
D_{m}^{\rm off} \equiv \int d\lambda \rho_{m}(\lambda) D_{\lambda}
\ee
where 
$
\rho_{m}(\lambda)
$
is some function converging, say for 
$
\lambda \rightarrow 0
$
to the distribution 
$
\delta(\lambda - m).
$
In this way all the fields from Section \ref{free} we become {\it generalized free fields} \cite{J} i.e. they will verify all properties described there except Klein-Gordon and Dirac equations. For instance the off-shell scalar field we be denoted by
$
\Phi^{\rm off}
$,
etc. However, for simplicity we will skip the index {\it off} if it is obvious from the context if we are considering on-shell or off-shell fields.
 
If we keep the definion of the gauge charge unchanged we will loose the property
$
Q^{2} = 0.
$ 
If we keep unchanged the expressions of the interaction Lagrangians from the preceding Section, but replace all fields by their off-shell counterparts, we also loose the relations (\ref{descent1}). However, these relations will be replaced by 
\be
d_{Q}T^{I} = i~\partial_{\mu}T^{I\mu} + S^{I}
\label{descent1-off}
\ee
with 
$
S^{I}
$
some polynomials which will be null in the on-shell limit. We will need these expressions in the following. In the following we will denote
$
K_{c} \equiv K_{m_{c}}
$
and we assume that all fields are off-shell (we do not append the index {\it off}). We have by direct computations the following results.

A. {\bf The Yang-Mills Sector} In this sector we have
\begin{thm}
The expressions
$
S_{ym}^{I} 
$
have the following explicit form:
\bea
S_{ym} = S_{ym}^{\emptyset} \equiv i~f_{abc}~u_{a}~\left(v_{b}^{\mu}~K_{c}v_{c\mu}
+ {1\over 2}~u_{a}~u_{b}~K_{c}\tilde{u}_{c}\right)
- i~f^{\prime}_{abc}~u_{a}~\Phi_{b}~K_{c}\Phi_{c}
\nonumber \\
- i~u_{a}~\d_{\mu}\bar{\Psi}~t_{a}^{\epsilon} \otimes \gamma^{\mu}~\gamma_{\epsilon}~\Psi
- ~u_{a}~\bar{\Psi}~M~t_{a}^{\epsilon} \otimes \gamma^{\mu}~\gamma_{\epsilon}~\Psi
\nonumber \\
- i~u_{a}~\bar{\Psi}~t_{a}^{\epsilon} \otimes \gamma^{\mu}~\gamma_{\epsilon}~\d_{\mu}\Psi
+ u_{a}~\bar{\Psi}~t_{a}^{- \epsilon}~M \otimes \gamma^{\mu}~\gamma_{\epsilon}~\Psi
\eea
Also
\be
S^{\mu} \equiv {i\over 2}~f_{abc}~u_{a}~u_{b}~K_{c}v_{c}^{\mu}
\ee
and
\be
S^{I} = 0,\quad |I| > 1. 
\ee
\label{ym-s}
\end{thm}
\newpage
B. {\bf The Gravitational Sector} In this case we have
\begin{thm}
The expressions
$
S^{I} 
$
have the following explicit form:
\bea
S = S^{\emptyset} \equiv i~\kappa~(2~u^{\mu}~\d_{\mu}h_{\alpha\beta}~K_{0}h^{\alpha\beta}
+ 2~\d_{\mu}u^{\mu}~h_{\alpha\beta}~K_{0}h^{\alpha\beta}
- u^{\mu}~\d_{\mu}h~K_{0}h
\nonumber \\
- \d_{\mu}u^{\mu}~h~K_{0}h
+ 2~\d_{\alpha}u_{\beta}~h^{\alpha\beta}~K_{0}h
- 4~\d^{\rho}u_{\mu}~h_{\nu\rho}~K_{0}h^{\mu\nu}
+ 2~u^{\mu}~\d_{\mu}u_{\nu}~K_{0}\tilde{u}^{\nu})
\eea
Also
\be
S^{\mu} \equiv - 2 i~\kappa~u^{\rho}~\d_{\rho}u_{\nu}~K_{0}h^{\mu\nu}
\ee
and
\be
S^{\mu\nu} = i~\kappa~( K_{0}u^{\nu}~u_{\rho}~\d^{\rho}u^{\mu} 
+ K_{0}u_{\lambda}~\d^{\lambda}u^{\mu}~u^{\nu}) - (\mu \leftrightarrow \nu)
\ee
\end{thm}

C. {\bf Interaction between Yang-Mills and Gravitation} In this case we have
\begin{thm}
The expressions
$
S^{I} 
$
have the following explicit form:
\bea
S = S^{\emptyset} \equiv 4 i~f_{ab}~( u^{\mu}~F_{a\mu\nu}~K_{b}v_{b}^{\nu}
- 4 i~u_{\mu}~K_{a}\Phi_{a}~\phi_{b}^{\mu})
+ i~f^{\prime}_{cd}~u^{\mu}~\d_{\mu}\Phi_{c}~K_{d}\Phi_{d}
\nonumber \\
+ {i\over 2}~u_{\mu}~
K\bar{\Psi} c^{\epsilon} \otimes \gamma^{\mu} \gamma_{\epsilon}~\Psi
- {i\over 2}~u_{\mu}~
\bar{\Psi} c^{\epsilon} \otimes \gamma^{\mu} \gamma_{\epsilon}~K\Psi
\nonumber \\ 
+ {i\over 2}~u_{\mu}~
\d^{\mu}\d^{\nu}\bar{\Psi} c^{\epsilon} \otimes \gamma_{\nu} \gamma_{\epsilon}~\Psi
+ {1\over 2}~u_{\mu}~
\d^{\mu}\bar{\Psi} M~c^{\epsilon} \otimes  \gamma_{\epsilon}~\Psi
\nonumber \\
+ {i\over 2}~u_{\mu}~
\d^{\mu}\bar{\Psi} c^{\epsilon} \otimes \gamma_{\nu}~\gamma_{\epsilon}~\d^{\nu}\Psi
- {1\over 2}~u_{\mu}~
\d^{\mu}\bar{\Psi} c^{- \epsilon}~M \otimes  \gamma_{\epsilon}~\Psi
\nonumber \\
- {i\over 2}~u_{\mu}~
\d^{\nu}\bar{\Psi} c^{\epsilon} \otimes \gamma_{\nu} \gamma_{\epsilon}~\d^{\mu}\Psi
- {1\over 2}~u_{\mu}~
\bar{\Psi} M~c^{\epsilon} \otimes  \gamma_{\epsilon}~\d^{\mu}\Psi
\nonumber \\
- {i\over 2}~u_{\mu}~
\bar{\Psi} c^{\epsilon} \otimes \gamma_{\nu} \gamma_{\epsilon}~\d^{\nu}\d^{\mu}\Psi
+ {1\over 2}~u_{\mu}~
\bar{\Psi} c^{- \epsilon}~M \otimes  \gamma_{\epsilon}~\d^{\mu}\Psi
\eea
and
\be
S^{I} = 0 \qquad |I| >1.
\ee
\end{thm}

Now we proceed to higher orders of perturbation theory and we have the following result. 
\begin{thm}
In the off-shell formalism we can choose the the second order chronological products such that the following identity is true:
\bea
d_{Q}T(T^{I_{1}}(x),T^{I_{2}}(y)) = i {\d \over \d x^{\mu}}T(T^{I_{1}\mu}(x),T^{I_{2}}(y))
+ (-1)^{|I_{1}|} {\d \over \d y^{\mu}}T(T^{I_{1}}(x),T^{I_{2}\mu}(y)
\nonumber \\
+ T(S^{I_{1}}(x),T^{I_{2}}(y)) + (-1)^{|I_{1}|}~T(T^{I_{1}}(x),S^{I_{2}}(y)).
\label{gauge-off}
\eea
\end{thm}
Indeed, if we make the substitution
$
D_{m}^{\rm off} \rightarrow D_{m}^{F,\rm off}
$
we obtain immediately the identity from (\ref{d-2}) and (\ref{descent1-off}). Similar identities are true in the higher orders of perturbation theory. Let us consider the simplest case 
$
I_{1} = I_{2} = \emptyset
$
when we have
\bea
d_{Q}T(T(x),T(y)) = i {\d \over \d x^{\mu}}T(T^{\mu}(x),T(y))
+ {\d \over \d y^{\mu}}T(T(x),T^{\mu}(y))
\nonumber \\
+ T(S(x),T(y)) + (x \leftrightarrow y).
\label{gauge-off-empty}
\eea
Now we have a very clear origin of the anomalies. It is elementary to prove that
\bea
T(S(x),T(y))^{\rm tree} = 
\sum_{m}~[~K_{m}D_{m}^{F,\rm off}(x - y)~A_{m}(x,y) 
+ \d_{\alpha}~K_{m}D_{m}^{F,\rm off}(x - y)~A^{\alpha}_{m}(x,y) 
\nonumber \\
+ \d_{\alpha}\d_{\beta}~K_{m}D_{m}^{F,\rm off}(x - y)~A^{\{\alpha\beta\}}_{m}(x,y)~]
+ \cdots
\eea
where by $\cdots$ we mean terms where the Klein-Gordon operator (or Dirac operator) is acting on some off-shell field factor. So when we make the on-shell limit 
$
\lambda \rightarrow 0
$
we have 
\be
T(S(x),T(y))^{\rm tree} \rightarrow  
~\delta(x - y)~A(x,y) + \d_{\alpha}\delta(x - y)~A^{\alpha}(x,y) 
+ \d_{\alpha}\d_{\beta}\delta(x - y)~A^{\{\alpha\beta\}}(x,y)
\ee
where the expressions
$
A(x,y), A^{\alpha}(x,y), A^{\{\alpha\beta\}}(x,y)
$
are sums of the corresponding expressions
$
A_{m}(x,y), A^{\alpha}_{m}(x,y), A^{\{\alpha\beta\}}_{m}(x,y).
$
In this way we have a systematic procedure to compute the tree anomalies in the second order of perturbation theory. For instance, the anomaly of the relation (\ref{gauge-off-empty}) is
\be
A(x,y) = \{\delta(x - y)~A(x,y) + [\d_{\alpha}\delta(x - y)]~A^{\alpha}(x,y) 
+ [\d_{\alpha}\d_{\beta}\delta(x - y)]~A^{\{\alpha\beta\}}(x,y)\}
+ (x \leftrightarrow y).
\label{ano-a}
\ee

We investigate now in what conditions we can eliminate the anomaly by finite renormalizations. The first trick is to use ``partial integration" on the last terms with derivatives on the $\delta$ distribution. We obtain the equivalent form:
\be
A(x,y) = 2~\delta(x - y)~a(x,y) 
+ \left[{\d \over \d x^{\alpha}}a^{\alpha}(x,y) + (x \leftrightarrow y) \right]
\label{ano-b}
\ee
where
\be
a(x,y) \equiv A(x,y) - {\d \over \d x^{\alpha}}A^{\alpha}(x,y)
+ {\d^{2} \over \d x^{\alpha}\d x^{\beta}}A^{\{\alpha\beta\}}(x,y)
\label{a}
\ee
and
\be
a^{\alpha}(x,y) \equiv [\d_{\beta}\delta(x - y)]~A^{\{\alpha\beta\}}(x,y) 
+ \delta(x - y)~
\left[A^{\alpha}(x,y) - {\d \over \d x^{\beta}}A^{\{\alpha\beta\}}(x,y)\right]  
\ee

If we make the redefinition
\be
T(T^{\mu}(x),T(y)) \rightarrow T(T^{\mu}(x),T(y)) + i~a^{\mu}(x,y)
\ee
of the chronological products we will put the anomaly in the form 
\be
A(x,y) = 2~\delta(x - y)~a(x,x) 
\label{ano-c}
\ee

Now we have the following 
\begin{lemma}
The preceding anomaly can be eliminated iff the expression
$
a(x) = a(x,x) 
$
is a relative cocycle i.e. we have
\be
a = d_{Q}B - i \d_{mu}B^{\mu}
\label{b}
\ee
for some Wick polynomials $B$ and
$
B^{\mu}
$.
The Wick polynomials
$
B(x)
$
and
$
B^{\mu}(x)
$
are constrained by: (a) Lorentz invariance; (b) ghost number restrictions:
\be
gh(B) = 0,\qquad gh(B^{\mu}) = 1
\ee
and (c) power counting which in our case gives:
\be
\omega(B)~,\omega(B^{\mu}) \leq 6.
\ee
\end{lemma}
The proof is very simple. Suppose that the anomaly (\ref{ano-c}) can be put in the form
\be
\delta(x - y)~a(x) =
d_{Q}R(x,y) + i {\d \over \d x^{\mu}}R^{\mu}(x,y) + {\d \over \d y^{\mu}}R^{\mu}(y,x).
\label{gauge-ano}
\ee 
with the expressions
$
R(x,y), R^{\mu}(x,y)
$
quasi-local i.e. of the form 
\be
R(x,y) = \delta(x - y)~B(x) + \cdots,\qquad
R^{\mu}(x,y) = \delta(x - y)~B^{\mu}(x)  + \cdots
\ee
where $\cdots$ are terms with higher order derivatives on the $\delta$ distribution. Then we immediately obtain from (\ref{gauge-ano}) the identity from the lemma. Conversely, if the identity from the lemma is true then we take
\be
R(x,y) = \delta(x - y)~B(x),\qquad
R^{\mu}(x,y) = \delta(x - y)~B^{\mu}(x)
\ee
and we have (\ref{gauge-ano}).

So all we have to do it to compute the expression
$
a(x,y)
$
given by the formula (\ref{a}), collapse the two variables to obtain the expression
$
a(x)
$
and impose the condition (\ref{b}). For simple models, as pure Yang-Mills theories, this computation is not very difficult but for more complicated models involving gravitation, the computation are very long and one can see the benefits of the off-shell method if one makes the comparison with the usual methods.

In the same way one can treat the other identities of the type (\ref{gauge-off}) i.e. for non-trivial sets of indices
$
I_{1},~I_{2}
$.
\newpage
\section{Second Order Gauge Invariance\label{second}}

Now we turn to the question of gauge invariance of the model in the second order of perturbation theory. The case of Yang-Mills fields has been investigated previously \cite{standard}. One can eliminate the ``tree" anomalies from the second order of the perturbation theory {\it iff} the following identities are true:
\be
f_{abc}f_{dec} + f_{bdc} f_{aec} + f_{dac} f_{bec} = 0, 
\label{Jacobi}
\ee
\be
f^{\prime}_{dca} f^{\prime}_{ceb} - f^{\prime}_{dcb} f^{\prime}_{cea} 
= - f_{abc} f^{\prime}_{dec},
\label{repr-f'}
\ee
\be
f^{\prime}_{cab} f^{\prime\prime}_{cde} + f^{\prime}_{cdb} f^{\prime\prime}_{cae} + f^{\prime}_{ceb} f^{\prime\prime}_{cda} = 0,
\quad \forall~b \in I_{1},
\label{h3-0}
\ee 
\be
[t_{a}^{\epsilon},t_{b}^{\epsilon}] = i f_{abc} t_{c}^{\epsilon}, 
\label{repr}
\ee
\be
t_{a}^{-} s_{b}^{+} - s_{b}^{+} t_{a}^{+} = i f'_{bca} s_{c}^{+}.
\label{WE}
\ee

For the case of pure gravity one can prove that the second order anomalies can be eliminated without further constraints \cite{Sc2}. Using the off-shell method one can compute the anomaly in a more systematic way. The anomaly can be eliminated without further constraints even in the case of massive gravity \cite{massive}. It remains to analyze the part of the interaction Lagrangian which describes the interaction between gravity and the other fields. The relevant expressions are 
$
T^{I}_{\rm int}, S^{I}_{\rm int}
$ 
given previously. The main result is 
\begin{thm}
The second order chronological products verify the gauge invariance condition 
\be
d_{Q}T(t_{\rm int}(x),t_{\rm int}(y)) = 
i {\d \over \d x^{\mu}}T(t_{\rm int}^{\mu}(x),t_{\rm int}(y))
+ {\d \over \d y^{\mu}}T(t_{\rm int}(x),t_{\rm int}^{\mu}(y))
\label{t-2=int}
\ee
in the second order of perturbation theory iff:
\bea
f_{ab} = - {\kappa \over 2}~\delta_{ab},~~a, b \in I_{1} \cup I_{2}
\nonumber \\
f^{\prime}_{cd} = 2~\kappa~\delta_{cd},~~c,d \in I_{3}
\nonumber \\
c^{\epsilon} = - i~\kappa
\eea 
The finite renormalization of the chronological product
$
T(t_{\rm int}(x),t_{\rm int}(y))
$
is given by the expression
$
R(x,y) = 2~\delta(x - y)~B(x)
$
where
\bea
B = - {i\over 2}~\kappa^{2}~(h_{\mu\nu}~h^{\mu\nu}~F_{a}^{\rho\sigma}~F_{a\rho\sigma}
+ h^{2}~F_{a}^{\rho\sigma}~F_{a\rho\sigma}
- 8~h_{\mu\nu}~h_{\rho\sigma}~F_{a}^{\mu\rho}~F_{a}^{\nu\sigma}
+ 8~h_{\mu\nu}~h~F_{a}^{\mu\rho}~F^{\nu}_{a\sigma})
\nonumber \\
+ 2~i~\kappa^{2}~\left( h_{\mu\nu}~h^{\mu\nu}~\Phi_{a}~\Phi_{a}
- {1\over 2}~h^{2}~\Phi_{a}~\Phi_{a}\right)
\nonumber \\
- i\kappa~f_{abc}~( - h~v_{a\mu}~v_{b\nu}~F_{c}^{\mu\nu}
+ 4 v_{a\mu}~v_{b\nu}~h^{\mu\rho}~F^{\nu}_{b\rho}
+ 2 u_{\mu}~v_{a}^{\mu}~v_{b}^{\nu}~\d_{\nu}\tilde{u}_{c})
\nonumber \\
- 8 i \kappa~f_{bca}~( 2 h_{\mu\nu}~v_{a}^{\mu}~\Phi_{b}~\d^{\nu}\Phi_{c}
+ 2 m_{b}~h_{\mu\nu}~v_{a}^{\mu}~v_{b}^{\nu}~\Phi_{c}
+ m_{b}~u_{\mu}~\tilde{u}_{a}~v_{b}^{\mu}~\Phi_{c})
\nonumber \\
+ i \kappa~f^{\prime\prime}_{abc}~h~\Phi_{a}~\Phi_{b}~\Phi_{c}
\nonumber \\
- 2i~\kappa^{2}~\Bigl[ \left( h_{\mu\nu}~h^{\mu\nu} - {1\over 4}~ h^{2}\right)~ 
\bar{\psi}~M~\otimes ~\gamma_{\epsilon} \psi
\nonumber \\
+ i~u_{\mu}~\tilde{u}_{\nu}~
(\d^{\mu}\bar{\psi}~\gamma^{\nu}\gamma_{\epsilon} \psi
+ \d^{\nu}\bar{\psi}~\gamma^{\mu}\gamma_{\epsilon} \psi
- \bar{\psi}~\gamma^{\mu}\gamma_{\epsilon} \d^{\nu}\psi
- \bar{\psi}~\gamma^{\nu}\gamma_{\epsilon} \d^{\mu}\psi)
\nonumber \\
+ {i \over 4}~u_{[\mu\nu]}~\tilde{u}_{\rho}~
\bar{\psi}~\gamma^{[\mu\nu\rho]}\gamma_{\epsilon} \psi
+ {i \over 2}~(\d_{\mu}h_{\nu\sigma}~{h^{\sigma}}_{\cdot\rho} 
- \d_{\mu}u_{\nu}~\tilde{u}_{\rho})~
\bar{\psi}~\gamma^{[\mu\nu\rho]}\gamma_{\epsilon} \psi
\nonumber \\
- {i \over 2}~({h_{\mu}}^{\rho}~h_{\nu\rho} + h~h_{\mu\nu} - 4 u_{\mu}~\tilde{u}_{\nu})~
(\d^{\mu}\bar{\psi}~\gamma^{\nu}~\gamma_{\epsilon} \psi
- \bar{\psi}~\gamma^{\nu}~\gamma_{\epsilon} \d^{\mu}\psi)\Bigl]
\nonumber \\
+ 2i~\kappa~v_{a}^{\mu}~h_{\mu\nu}~
\bar{\psi}~t^{\epsilon}_{a}~\otimes ~\gamma^{\nu}~\gamma_{\epsilon} \psi
\eea
\end{thm}
{\bf Proof:} (i) We will compute the anomaly using the off-shell method described in the preceding Section. We first consider the contribution to the anomaly without Dirac fields. 
The expression is very long (it has $75$ terms!) and we observe that we can consider independently the contribution of the type
$
x_{1}~x_{2}~y_{1}~y_{2}
$
and
$
x~y_{1}~y_{2}~y_{3}
$
where
$
x_{j}
$
are gravitational variables and
$
y_{j}
$
are Yang-Mills variables. So the expression $a$ of the anomaly has two contributions of the two types described above. After a long computation one can put the first contribution in the form:
\be
a_{1} = d_{Q}B_{1} + {\rm total~divergence} 
+ 2~g_{ab}~W_{ab} - 2~g^{\prime}_{ab}~W^{\prime}_{ab} + \cdots
\label{a1}
\ee
where the matrices
$
g, g^{\prime}
$
are given by
\bea
g \equiv 2 f^{2} + \kappa~f
\nonumber \\
g^{\prime} \equiv 16 f^{2} + 8~\kappa~f + (f^{\prime})^{2} -2 \kappa~f^{\prime}
\label{c}
\eea
and the Wick polynomials
$
W_{ab},~W^{\prime}_{ab}
$
by
\bea
W_{ab} \equiv 4 u^{\mu}~\d_{\mu}u_{\nu}~F_{a}^{\nu\rho}~\d_{\rho}\tilde{u}_{b}
- 8~\d^{\rho}u^{\mu}~h_{\nu\rho}~F_{a}^{\mu\lambda}~F_{b\lambda}^{\nu}
+ 4~\d_{\mu}u^{\mu}~h_{\rho\sigma}~F_{a}^{\rho\lambda}~F_{b\lambda}^{\sigma}
\nonumber \\
+ 4~u^{\mu}~\d_{\mu}h_{\rho\sigma}~F_{a}^{\rho\lambda}~F_{b\lambda}^{\sigma}
- u_{\mu}~\d_{\mu}h~F_{a}^{\rho\sigma}~F_{b\rho\sigma}
\eea
and 
\be
W^{\prime}_{ab} \equiv u^{\rho}~(\d_{\mu}h^{\mu\nu}~\d_{\rho}\Phi_{a}~\d_{\nu}\Phi_{b}
+ h^{\mu\nu}~\d_{\rho}\Phi_{a}~\d_{\mu}\d_{\nu}\Phi_{b}).
\ee
We have denoted by $\cdots$ in (\ref{a1}) the terms of canonical dimension less than $6$ (these terms are monomials with mass factors). The elimination of the anomaly (\ref{a1}) is possible {\it iff} we have
\be
g_{ab}~W_{ab} - g^{\prime}_{ab}~W^{\prime}_{ab} 
= d_{Q}B^{\prime}_{1} + {\rm total~divergence}
\ee
From here we obtain
\be
g_{ab}~d_{Q}W_{ab} - g^{\prime}_{ab}~d_{Q}W^{\prime}_{ab} = {\rm total~divergence}
\ee
and if we make a generic ansatz for the divergence in the right hand side we can prove that we have
$
g = 0,~g^{\prime} = 0
$
i.e.
\be
2 f^{2} + \kappa~f = 0,
\qquad
(f^{\prime})^{2} -2 \kappa~f^{\prime} = 0.
\label{f}
\ee

One can use the preceding relation to simplify considerably the forms of the matrices 
$
f,~f^{\prime}
$.
Indeed, we already know that these matrices are real and symmetric, so they can by diagonalized with orthogonal matrices:
\be
f = C~F~C^{-1},\qquad f^{\prime} = C^{\prime}~F^{\prime}~(C^{\prime})^{-1}
\ee
where
\be
F_{ab} = F_{a}~\delta_{ab},\qquad F^{\prime}_{ab} = F^{\prime}_{a}~\delta_{ab}.
\ee

From (\ref{f}) we have immediately
\be
2 F^{2} + \kappa~F = 0,
\qquad
(F^{\prime})^{2} -2 \kappa~F^{\prime} = 0.
\ee
and if we substitute the preceding diagonal expressions we obtain 
\be
F_{a} = - {\kappa \over 2},\qquad F^{\prime}_{a} = 2~\kappa,~~\forall a
\ee
and we obtain the diagonal form from the statement. Moreover, if we compute all the terms of 
$
a_{1}
$
(i.e. those with mass factors also) we obtain that we have
\be
a_{1} = d_{Q}B_{1} + {\rm total~divergence} 
\ee
with 
$
B_{1} 
$
given by the first two lines from the expression $B$ from the statement.

(ii) We turn now to the contribution
$
a_{2}
$
and after a long computation we obtain
\be
a_{2} = d_{Q}B_{2} + {\rm total~divergence} 
+ g^{(1)}_{abc}~W_{abc} - {1\over 2}~(g^{(2)}_{abc} + g^{(2)}_{bac})~W^{(2)}_{abc} 
+ g^{(3)}_{abc}~W^{(3)}_{abc} + \cdots
\label{a2}
\ee
where the constants $g$ are
\bea
g^{(1)}_{abc} \equiv f_{abc}~f_{cd} + (b \leftrightarrow c)
\nonumber \\
g^{(2)}_{abc} \equiv f^{\prime}_{cd}~f^{\prime}_{dba} - 4~f_{cd}~f^{\prime}_{dba}
\nonumber \\
g^{(3)}_{abc} \equiv g^{(2)}_{abc} - 4~f_{ad}~f^{\prime}_{bcd}
\eea
and the Wick monomials $W$ by
\bea
W^{(1)}_{abc} \equiv - 4 u_{\mu}~v_{a}^{\rho}~F_{b}^{\mu\nu}~F_{c\rho\nu}
+ u_{\mu}~v_{a}^{\rho}~F_{b}^{\rho\sigma}~F_{c\rho\sigma}
- 4 h_{\mu\nu}~u_{a}~F_{b}^{\mu\rho}~F_{c\rho}^{\nu}
\nonumber \\
+ h~u_{a}~F_{b}^{\mu\nu}~F_{c\mu\nu}
+ 4 u_{\mu}~u_{a}~F_{b}^{\mu\nu}~\d_{\nu}\tilde{u}_{c} 
\eea
\be
W^{(2)}_{abc} \equiv 2~u^{\mu}~v_{a}^{\nu}~\d_{\mu}\Phi_{b}~\d_{\nu}\Phi_{c}
- u_{\mu}~v_{a}^{\mu}~\d_{\nu}\Phi_{b}~\d^{\nu}\Phi_{c}
+ 2~h^{\mu\nu}~u_{a}~\d_{\mu}\Phi_{b}~\d_{\nu}\Phi_{c}
\ee
\be
W^{(3)}_{abc} \equiv u_{\mu}~F_{a}^{\mu\nu}~\Phi_{b}~\d_{\nu}\Phi_{c}
\ee
and the dots $\cdots$ are terms of canonical dimension less than $6$. However it is a remarkable fact that the expressions 
$
g^{(1)}_{abc},~g^{(2)}_{abc} + g^{(2)}_{bac}
$ 
and
$
g^{(3)}_{abc}
$
all null! This can be proved if we combine the diagonal form of the matrices  
$
f,f^{\prime}
$
with the other properties of the constants
$
f_{abc},~f^{\prime}_{abc}
$
obtained in the analysis of the Yang-Mills model. Moreover if we compute the mass-dependent contribution in the expression
$
a_{2}
$
we have after a long computation
\be
a_{2} = d_{Q}B_{2} + {\rm total~divergence} 
+ g_{abc}~u_{\mu}~\Phi_{a}~\Phi_{b}~\d^{\mu}\Phi_{c}
\label{a22}
\ee
where
\be
g_{abc} \equiv {1\over 2}~( - 4 f_{cd} + f^{\prime}_{cd})~f^{\prime\prime}_{dab}.
\ee
However, if we use as above the various properties of the constants
$
f,f^{\prime}
$
and
$
f^{\prime\prime}_{abc}
$
already obtained we can prove that the expression
$
g_{abc}
$
is completely symmetric so we easily show that the last term from (\ref{a22}) is a relative coboundary. We obtain that 
$
a_{2}
$ 
has the form 
\be
a_{2} = d_{Q}B_{2} + {\rm total~divergence} 
\ee
with 
$
B_{2}
$
given by the lines $3,4$ and $5$ of the expression $B$ from the statement.

(iii) We consider now the Dirac terms of the anomaly. Again we have a very long list ($73$ terms!) which can be group in two contributions
$
a_{3} \sim x_{1}~x_{2}~\bar{\psi}~\psi
$
and
$
a_{4} \sim x~y~\bar{\psi}~\psi
$
where as above 
$
x_{j}
$
are gravitational variables and
$
y_{j}
$
are Yang-Mills variables. We start with the contribution 
$
a_{3}
$
and after a long computation we can put it in the form 
\be
a_{3} = d_{Q}B_{3} + {\rm total~divergence} 
+ 2~u_{\mu}~h_{\rho\sigma}~A^{\mu\{\rho\sigma\}} + 2~i~u_{\mu}~h~A^{\mu}
\ee
where
\bea
A^{\mu\{\rho\sigma\}} = 
( \d^{\mu}\bar{\psi}~A^{\epsilon}~\otimes \gamma_{\sigma}~\gamma_{\epsilon} \d^{\rho}\psi
- \d^{\rho}\bar{\psi}~A^{\epsilon}~\otimes \gamma_{\sigma}~\gamma_{\epsilon} \d^{\mu}\psi)
+ ( \rho \leftrightarrow \sigma)
\nonumber \\
+ \d^{\rho}\d^{\sigma}\bar{\psi}~A^{\epsilon}~\otimes \gamma_{\mu}~\gamma_{\epsilon} \psi
- \bar{\psi}~A^{\epsilon}~\otimes \gamma_{\mu}~\gamma_{\epsilon} \d^{\rho}\d^{\sigma}\psi
\eea
\be
A^{\mu} = \d^{\mu}\bar{\psi}~M~A^{\epsilon}~\otimes ~\gamma_{\epsilon} \psi
+ \bar{\psi}~M~A^{\epsilon}~\otimes ~\gamma_{\epsilon} \d^{\mu}\psi
- \bar{\psi}~A^{\epsilon}~\otimes \gamma_{\mu}~\gamma_{\epsilon} \d^{\rho}\d^{\sigma}\psi
\ee
and we have defined the matrices
\be
A^{\epsilon} \equiv \kappa~c^{\epsilon} - i~(c^{\epsilon})^{2},\qquad \epsilon = \pm.
\ee

We must have 
\be
u_{\mu}~h_{\rho\sigma}~A^{\mu\{\rho\sigma\}} + ~i~u_{\mu}~h~A^{\mu}
= d_{Q}B - i \d_{\mu}B^{\mu}
\ee
and if we make an ansatz for the relative coboundary from the right hand side we immediately obtain that we must have
$
A^{\mu\{\rho\sigma\}} = 0,~A^{\mu} = 0
$
and this leads to
$
A^{\epsilon} = 0.
$
i.e.
\be
(c^{\epsilon})^{2} = - i~\kappa~c^{\epsilon},\qquad \forall \epsilon = \pm.
\label{C}
\ee

We can use this relation to simplify the expressions for the matrices
$
c^{\epsilon}
$.
Indeed, we have previously derived the self-adjointness property
$
(c^{+})^{\dagger} = - c^{-}
$
so the matrices 
\be
c \equiv {1 \over 2}~(c^{+} + c^{-})
\qquad
c^{\prime} \equiv {1 \over 2}~(c^{+} - c^{-})
\ee
are verifying
\be
c^{\dagger} = - c,\qquad (c^{\prime})^{\dagger} = c^{\prime}.
\ee

Because the matrix $c$ is anti-self-adjoint we can diagonalize it with a unitary transformation:
$
c = U~C~U^{-1}
$
with 
$
C_{jk} = \delta_{jk}~f_{j}.
$
Similarly we define 
$
c^{\prime} = U~C^{\prime}~U^{-1}
$
so we have
$
c^{\epsilon} = U~C^{\epsilon}~U^{-1}.
$
Now we have from  relation (\ref{C}) that 
\bea
(C^{\epsilon})^{2} = - i~\kappa~C^{\epsilon},\qquad \forall \epsilon = \pm 
\qquad  \Longleftrightarrow
\nonumber \\
C^{2} + (C^{\prime})^{2} = - i~\kappa~C,
\qquad
C~C^{\prime} + C^{\prime}~C = i~\kappa~C^{\prime}
\label{C2}
\eea
and this leads to the conclusion that the matrix
$
(C^{\prime})^{2}
$
is diagonal. Because all these matrices are finite dimensional, this means that the matrix
$
C^{\prime}
$
is also diagonal. If we substitute in (\ref{C2}) we get that the matrices
$
C^{\epsilon}
$ 
are in fact proportional to the unit matrix. This fact is true then for the original
matrices
$
c^{\epsilon}
$
also. If we insert
$
c^{\epsilon} = \lambda^{\epsilon}~{\rm Id},~\lambda^{\epsilon} \in \C
$
in (\ref{C}) we get 
$
\lambda^{\epsilon} = - i~\kappa,\forall \epsilon
$
and this leads to the expression for 
$
c^{\epsilon}
$
from the statement. This condition ensures that
\be
a_{3} = d_{Q}B_{3} + {\rm total~divergence}
\ee
where 
$
B_{3}
$
are the lines $6,7$ and $8$ of the expression $B$ from the statement of the theorem.

(iv) Now we still have to analyze the contribution
$
a_{4}.
$
After a long computation we get
\be
a_{4} = d_{Q}B_{4} + {\rm total~divergence}
\ee
where 
$
B_{4}
$
is the last line of the expression $B$ from the statement of the theorem.

The expression for the counterterm
$
B^{\mu}
$
follows also from the preceding computations but we skip it.
$\qed$

%\newpage
\section{Conclusions}

Second order gauge invariance does not lead to some new information concerning the Yang-Mills sector. We only get the diagonal structure of the matrices
$
f,~f^{\prime}
$
and
$
c^{\epsilon}
$
for the interaction sector. This result can be interpreted as the {\it universality} of the interaction of gravity with other fields: one does not need new constraints to make the existence of the interaction Lagrangian possible. The extension of the method to the case of massive gravity is elementary and leads to the results from \cite{mass+gravity}. 

In further publications we will consider loop contributions in the second order of the perturbation theory and also higher order of perturbation theory using the same ideas.

\end{document}